\documentclass{agujournal2019}
\usepackage{url} 
\usepackage{lineno}
\usepackage{soul}
\usepackage{natbib}

\draftfalse

\journalname{}

\begin{document}

\title{Satellites and small bodies with ALMA: Insights into Solar System formation \& evolution}

\authors{Katherine de Kleer\affil{1}, Michael E. Brown\affil{1}, Martin Cordiner\affil{2,3}, Richard Teague\affil{4}}

\affiliation{1}{Division of Geological and Planetary Sciences, Caltech. 1200 E California Blvd, Pasadena CA 91125 USA.}
\affiliation{2}{Astrochemistry Laboratory, NASA Goddard Space Flight Center, 8800 Greenbelt Road, Greenbelt, MD 20771, USA.}
\affiliation{3}{Department of Physics, Catholic University of America, Washington, DC 20064, USA.}
\affiliation{4}{Department of Earth, Atmospheric, and Planetary Sciences, Massachusetts Institute of Technology, Cambridge, MA 02139, USA}

\correspondingauthor{Katherine de Kleer}{dekleer@caltech.edu}

\begin{keypoints}
\item Discoveries from ALMA have led to progress in our understanding of planet formation within our solar system and in exoplanetary systems.
\item ALMA resolves thermal emission from debris in disks, and from the remnants of that debris in our own Solar System -- asteroids and KBO’s.
\item ALMA maps the same gas species in planets, moons, and comets as in protoplanetary disks, enabling chemical inheritance studies.
\end{keypoints}

\begin{abstract}

Our understanding of the formation and evolution of planetary systems has made major advances in the past decade. This progress has been driven in large part by the Atacama Large Millimeter/submillimeter Array (ALMA), which has given us an unprecedented view of Solar System bodies themselves, and of the structure and chemistry of forming exoplanetary systems. Within our own Solar System, ALMA has enabled the detection of new molecules and isotopologues across moons and comets, as well as placing new constraints on the compositions and histories of small bodies through thermal emission observations. In this article, we highlight some key areas where ALMA has contributed to a deeper understanding of our Solar System's formation and evolution, and place these discoveries in the context of our evolving understanding of protoplanetary disks.

\end{abstract}

\section{Introduction}

The Atacama Large Millimeter/submillimeter Array (ALMA) has been transforming our understanding of planet formation since the first high-resolution images it returned of the disk around HL Tau in 2014 \citep{HLTau2015}. Observations of thermal emission from dust in protoplanetary disks have revealed widespread gap, ring, and spiral structures \citep{andrews2018,long2018}, motivating models for planet formation in ring structures including within our own Solar System \citep{Morbidelli2020,izidoro2022}. Spectral line observations, which are sensitive to numerous rotational transitions of gas-phase molecules, have mapped volatile gas species and their isotopologues in the outer disk regions (typically 10's to 100's of AU from the star), elucidating the origins and evolution of disk chemical composition \citep{oberg2021}. 

The structure and chemistry of our own protoplanetary disk set the early conditions for the suite of planets, moons, and small bodies we have in the Solar System today. Comparisons between these objects and their precursors, as observed in the disks of exoplanetary systems, help us understand the early evolution of the Solar System and place it in context within the diversity of forming planetary systems. Our Kuiper belt has its analog in exocometary belts \citep{matra2025}, and both the Kuiper and asteroid belts are the remains of planetesimal  populations analogous to those that collisionally generate the dusty belts of debris disks. ALMA's high sensitivity and angular resolution allow us to study emission from dust in these debris disks, and these same capabilities allow us to map thermal emission from our asteroids and Kuiper Belt Objects (KBO's). Such data on small Solar System objects directly measure and map asteroid and KBO surface properties such as emissivity, which constrains composition and surface texture, and also enable sufficiently precise astrometry to infer the formation histories of binary objects from their orbits. Similarly, the spectral lines that trace the gas chemistry in disks can also reveal the chemistry of planets, moons, and comets in the Solar System. When the same molecular species are measured at Solar System planets and in disks at the stellar distances at which the planets are thought to have formed, information on chemical inheritance from disk through planet phase can be inferred.

This article is focused on the formation and long-term evolution of Solar System objects, and how our understanding of these processes has been advanced by ALMA. We highlight a few key case studies from both thermal emission of surfaces (Section \ref{sec:thermal}) and spectral line observations of gasses (Section \ref{sec:chemistry}), placing these results in the context of our evolving understanding of planet formation processes. We discuss the major limitations on these comparisons and the potential for near-future advancement in Section \ref{sec:outlook}.

\section{Thermal Emission: Dust, KBOs, and Asteroids}\label{sec:thermal}

\begin{figure}
    \centering
    \includegraphics[scale=.5]{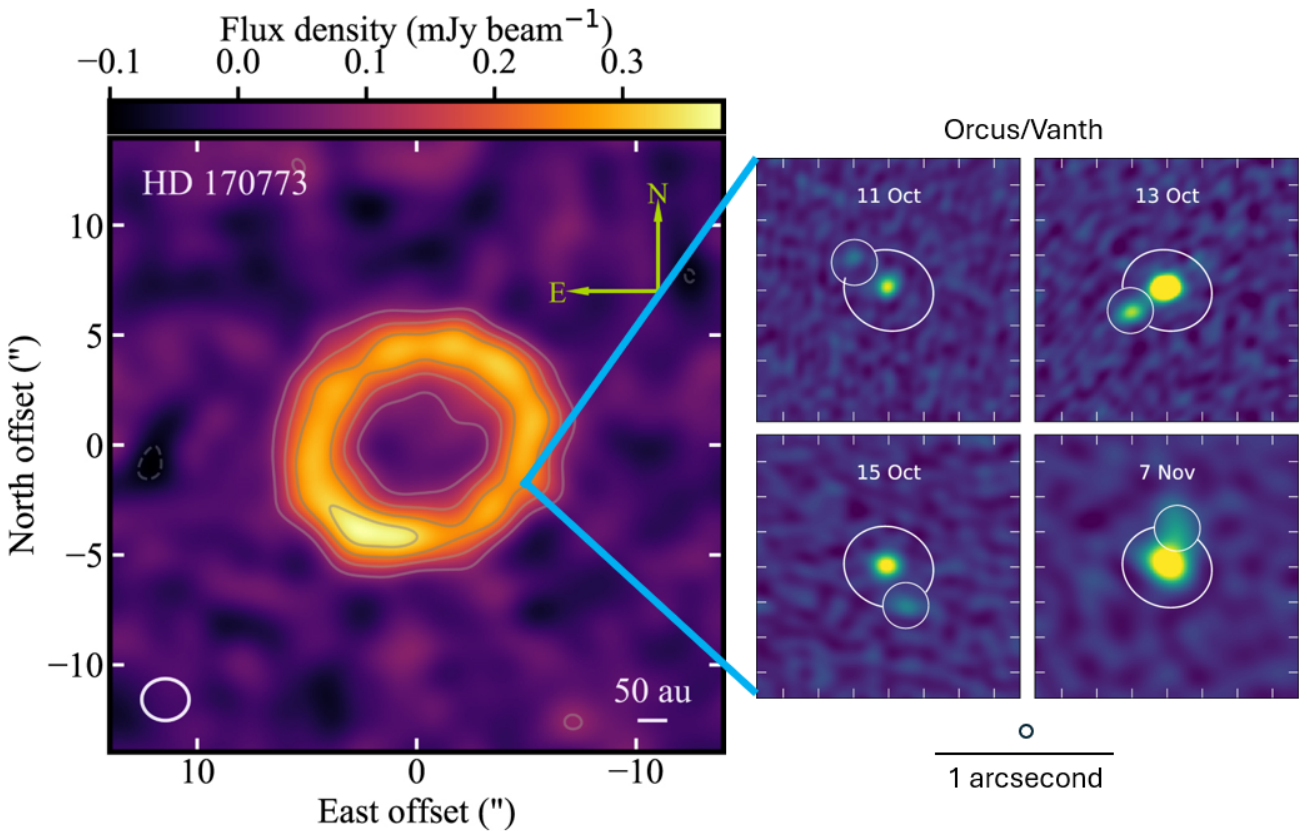}
    \caption{(Left panel) ALMA continuum thermal observations of the debris disk surrounding HD 170773, a nearby $\sim$1.3 solar mass star, from \cite{Sepulveda_2019}. The emission is from dust
    created from collisions within
    a large population of small bodies residing in
    the debris disk. (Right panel) A
    solar system example of the types of 
    bodies within the debris is the
    dwarf planet system Orcus and Vanth, from \cite{bb_size}.
    ALMA has detected the tiny barycentric wobble induced on Orcus by Vanth, with a size approximately
    equal to the circle above the ``1 arcsecond'' scale bar, showing that the mass ratio
    of this system is the highest of any measured primary-satellite
    in the solar system.}
    \label{fig:thermal}
\end{figure}

Thermal emission observations with ALMA allow an almost direct connection between nearby mature planetary systems and our own. Around nearby stars, thermal emission from disks of cold dust are detected in $\sim$20\% of systems \citep{Montesinos_2016,matra2025}. This dust – in what are called ``debris disks’’ – is the product of collisions between small bodies in the exo-planetary equivalent of our solar system’s Kuiper belt. Figure \ref{fig:thermal} shows one such ALMA detection of a debris disk around a nearby star with a mass about 1.3 times that of the sun \citep{Sepulveda_2019}. Such images can be used to trace the size and location of planetesimal belts, the influence of giant planets, and the dynamical evolution of the system.

In the solar system, the same ALMA characteristics that make it a powerful tool to study debris disks – namely the large collecting area and the high spatial resolution – have allowed direct study of the types of objects that are colliding in debris disks to produce the detected dust. Observations have been made of KBOs and their satellites \citep{lellouch,bb_size}, of small Jupiter Trojan asteroids \citep{anna}, and of asteroids in the main belt \citep{ast0,dekleer2021,dekleer2024_kalliope}. Such studies have been used to understand sizes and masses of bodies, properties of their surfaces and interiors, and their formation and evolution.

An instructive example of the power of ALMA observations in the outer solar system comes from the study of KBO-satellite systems. Depending on the collisional and accretionary environment of the early outer solar system, satellites can be primordial and form simultaneously with their primary \citep{nesvorny}, satellites can be captured with the gravitational assistance of other bodies in a dense early disk \citep{goldreich}, or they can be either largely intact or completely reaccreted products of a near-catastrophic giant impact \citep{canup} (of the sort that leads to the type of dust observable as a debris disk around other stars). Each of these scenarios produces a characteristic mass and density ratio between the satellite and the primary. Good examples of primordial formation include the small near-equal-sized and identically-colored cold classical KBO binaries \citep{noll}, while the small low density icy satellites of high density dwarf planet Haumea demonstrate the effects of reaccretion after a giant impact \citep{darin}. The moderate mass ratio Pluto-Charon system more closely resembles a capture, though one that was aided by energy dissipation in a grazing impact \citep{canup}. 

Measuring the mass ratio between the primary and the satellite is difficult without special circumstances like multiple satellites, but ALMA’s high spatial resolution (down to 20~mas) and its exquisite ability to measure an absolute position on the sky with-respect-to background quasars uniquely allows it to track the minute barycentric wobble of a KBO as it is perturbed by an orbiting satellite. \cite{BB_mass} exploited this capability to measure the densities of Vanth and Dysnomia, the satellites of the dwarf planets Orcus and Eris, respectively. The Vanth-Orcus barycentric motion is easily observable and reveals a mass ratio of 0.16 $\pm$0.02, the highest of any known planet or dwarf planet (Fig. \ref{fig:thermal}). This large ratio is consistent with the hypothesis that Vanth is a largely intact impactor from a giant collision in the system and that the system has likely evolved to a double synchronous state. No barycentric motion was detected at Eris, leading to a 0.0085 upper limit to the mass ratio, close to the ratio expected for a reaccreted satellite, yet Dysnomia was also found to have an albedo of only 0.05$\pm$0.01, wholly inconsistent with being a product of ice reaccretion.  Intriguingly, the low mass of Dysnomia, coupled with the observation that the rotation of Eris is locked to the 16 day orbital period of Dysnomia, requires the interior of Eris to be highly dissipative and thus that the ice shell is convective, in contrast to the conductive shell of Pluto, possibly as a result of impact-induced depletion of volatiles at Eris \citep{francis}. While the full implications of these observations are not yet clear, the power of ALMA is allowing the first geophysical characterization of dwarf planets at nearly 100 AU.

In addition to enabling precise astrometry, ALMA measurements of the thermal emission from a KBO or asteroid also provide information on the physical properties and composition of the regolith. KBO's, like other icy solar system bodies, exhibit low emissivities at millimeter wavelengths \citep[0.70$\pm$0.13;][]{lellouch2017}; while this can broadly be understood as a scattering effect and likely encodes information on grain size, there remain unexplained trends within the small sample size observed to-date. A larger sample size will be needed to determine how emissivity correlates with properties such as orbit and albedo and thus better understand what information it conveys about a surface. On rocky objects such as main belt asteroids, the metal content of the surface is the dominant influence on the emissivity, with higher metal content corresponding to lower emissivity \citep{parkhomenko1967,campbell1969}. As such, emissivity can be used to trace the products of planetesimal differentiation \citep{dekleer2021,cambioni2022}. ALMA's high angular resolution and sensitivity enable $\sim$30 km resolution within the main asteroid belt, providing 50--100 resolution elements over the disks of large asteroids. Although only a handful of main-belt asteroids have been observed in this way to date, the nominally metal-rich asteroids (`M-types') are shown to have emissivities around 0.60, while rocky asteroids (`S-types') are $>$0.75 \citep{dekleer2021,dekleer2024_kalliope,ast0,phua2024}. In cases of binary asteroids, the emissivities can be compared between the primary and secondary to differentiate between capture and collision formation scenarios. For example, Linus -- the satellite of asteroid (22) Kalliope -- was found to be metal depleted compared to Kalliope, while Kalliope's surface hosts a localized region of low emissivity / high metal content \citep{dekleer2024_kalliope}. This supports a formation scenario whereby the material that formed Linus was excavated from an impact onto an already-differentiated Kalliope.

\section{Chemistry of Volatiles}\label{sec:chemistry}

The elements carbon, hydrogen, oxygen, nitrogen, and sulfur (CHONS) are the primary volatile-forming elements in a planetary system. These elements form molecules that are found in the gas phase in the inner part of a protoplanetary disk, and in the ice phase out beyond each molecule's snow line. Which specific molecules are formed at a given stellar distance depends on the relative abundances of the elements and disk conditions including temperature and radiation field. Once incorporated into planetary-scale bodies, these elements make up the majority of the (sub)surface liquids and atmospheric gasses, and are central to the question of habitability.

A key question regarding these elements is to what extent the chemical inventory of a planet (including the specific molecular and isotopic make-up) is set by each of the prestellar, protostellar, disk, and planet phases. The primary avenue for investigating chemical inheritance through the latter stages observationally is to measure the ratios of atoms, molecules, and isotopes for a sample of objects at the disk and planet phases, at the same stellar distances.

Our Solar System provides the temporal end-member, which can be compared with forming planetary systems to trace inheritance. ALMA's wavelength coverage spans myriad rotational transitions of volatiles molecules, enabling abundance measurements of gas-phase species in both disks and Solar System bodies. Measurements of isotopic abundances give particularly powerful insight into the source region of an object's volatiles and how the volatile inventory has evolved over time due to processing and loss.

Although the most extensive isotopic measurements of individual Solar System objects come from \textit{in situ} spacecraft data \citep[e.g.][]{Altwegg2020,Nixon08}, Earth-based telescopes are able to survey a much larger range of objects. ALMA has provided measurements of the $^{13}$C/$^{12}$C, $^{17}$O/$^{16}$O, $^{18}$O/$^{16}$O, $^{15}$N/$^{14}$N and D/H ratios  at Titan \citep{serigano16,molter16,nosowitz25}, the $^{13}$C/$^{12}$C, $^{15}$N/$^{14}$N and D/H ratios in comets \citep{cordiner19,cordiner24,cordiner25}, and the $^{34}$S/$^{32}$S and $^{37}$Cl/$^{35}$Cl ratios at Io \citep{deKleer2024}. Here we focus on key contributions made by ALMA on using nitrogen and sulfur to investigate questions of planet formation and evolution. For a broader review on isotopes of CHON for understanding planet formation, see \cite{nomura23}.

\subsection{Nitrogen}

Nitrogen is a valuable tool for studying questions of formation and evolution because, unlike carbon and oxygen, its stable isotope ratios show large (orders of magnitude) variations between Solar System bodies. In protoplanetary disks, the higher column density of the dominant $^{14}$N$^{14}$N  isotopologue of N$_2$ compared with the less abundant $^{15}$N$^{14}$N isotopologue, leads to stronger self-shielding for $^{14}$N$^{14}$N, and therefore, $^{15}$N$^{14}$N is more rapidly photodissociated. This leads to a region of the disk becoming enriched in gas-phase $^{15}$N, which gets incorporated into other (gas and ice-phase) molecules, resulting in enhancements in their $^{15}$N/$^{14}$N ratios \citep{vis18,fur18}. The resulting enrichment in $^{15}$N has been observed in the HCN molecule in protoplanetary disks by ALMA, close to the location where planets and comets are forming around other stars \citep{Guzman2017,hily-blant2019,ber20}. 

\begin{figure}
\centering
\includegraphics[width=0.8\textwidth]{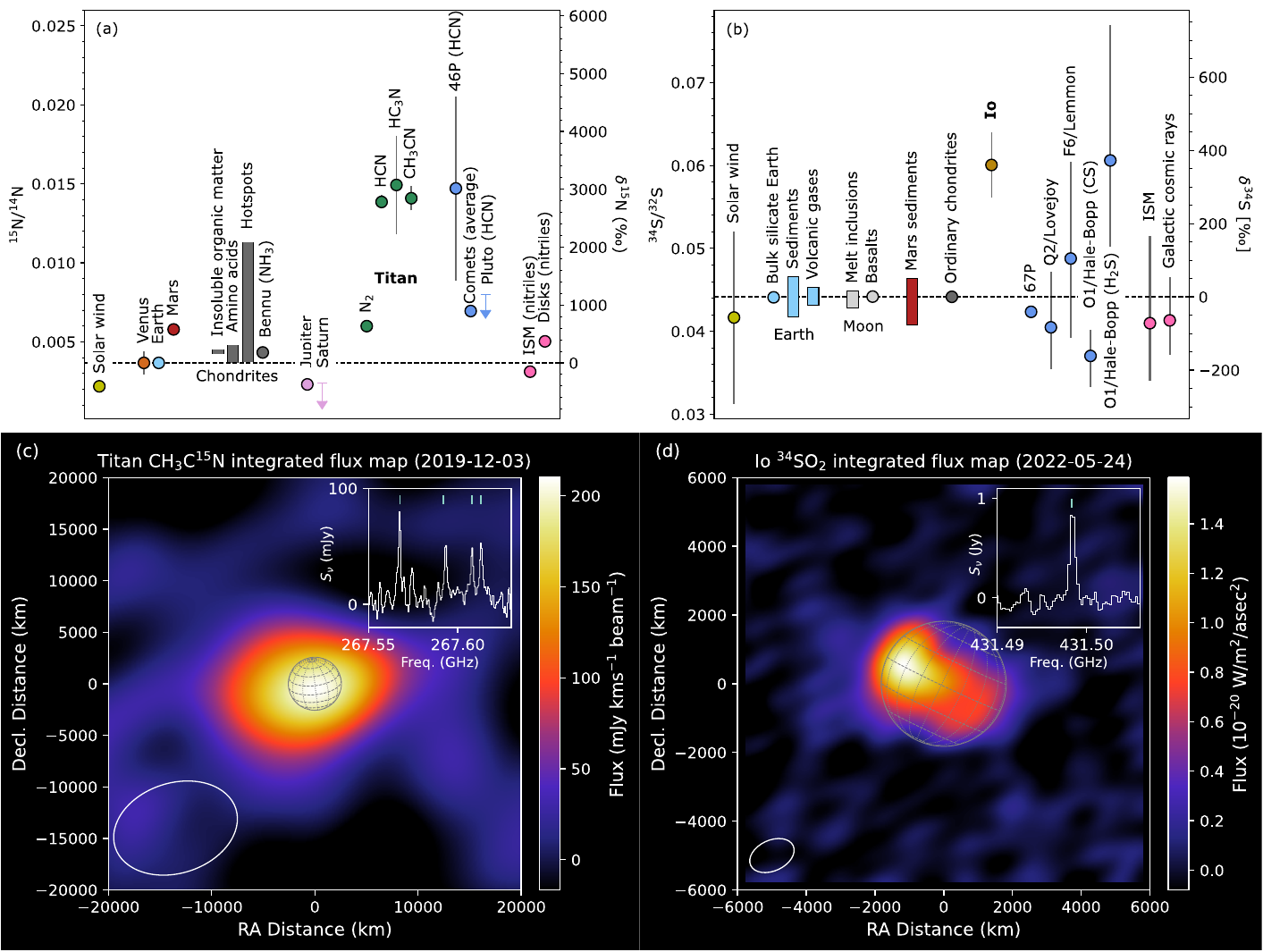}
\caption{(a) $^{15}$N/$^{14}$N ratios among various astronomical objects (from \citealt{lellouch2017_pluto,glavin2025,nosowitz25,cordiner24,nomura23}, and references therein). ALMA is responsible for the observations of HCN, HC$_3$N and CH$_3$CN on Titan, HCN in comet 46P/Wirtanen, the HCN upper limit at Pluto, and nitriles in protoplanetary disks. The average comets value is the error weighted mean of 31 measurements from \citet{hily17}. Horizontal dashed lines indicate the terrestrial values. (b) $^{34}$S/$^{32}$S ratios in the Solar System (fig adapted from \cite{deKleer2024}). Abbreviations are used for the designations of comets 67P/Churyumov–Gerasimenko, C/2014 Q2 (Lovejoy), C/2012 F6 (Lemmon), and C/1995 O1 (Hale-Bopp). (c) Map and spectrum of CH$_3$C$^{15}$N from ALMA used to derive the $^{15}$N/$^{14}$N ratio \citep{nosowitz25}, and (d) Map and spectrum of $^{34}$SO$_2$ from ALMA used to derive the $^{34}$S/$^{32}$S ratio \citep{deKleer2024}. Vertical ticks indicate the respective molecular line frequencies. \label{fig:15N}}
\end{figure}

The same process should leave an imprint of $^{15}$N enrichment on Solar System bodies that accreted from the icy parts of the disk. This is observed in the population of comets, which are ice-rich primordial bodies that accreted during the epoch of planet formation \citep{mumma2011}.
Figure \ref{fig:15N} shows $^{15}$N/$^{14}$N measurements in comets and other solar system bodies alongside observed values for the Sun, protoplanetary disks, and the interstellar medium (ISM). Comets possess a surprising degree of uniformity in their $^{15}$N/$^{14}$N ratios, among different molecules, and across different comets.
However, in 2018, ALMA observations of Jupiter family comet 46P/Wirtanen during its exceptionally close passage (within 0.1 au) of the Earth found a HC$^{15}$N/HC$^{14}$N ratio that was significantly higher than previously measured in N-bearing molecules in comets or other (bulk) Solar System bodies to-date \citep{cordiner24}. The $^{15}$N enrichment in comet 46P's HCN implies that ices with larger-than-normal $^{15}$N/$^{14}$N ratios became incorporated into this comet during accretion of its nucleus, which plausibly originated from part(s) of the protosolar disk where isotope-selective photodissociation was more efficient, and subsequent nitrile ice production occurred rapidly. Cometary $^{15}$N/$^{14}$N ratios may therefore be more diverse than previously thought, so additional studies are warranted, to help reveal the precise origin (and isotopic composition) of the materials from which comets were accreted.

At planets and moons, the chemical and physical evolution of their dynamic atmospheres causes the observed $^{15}$N/$^{14}$N ratios to change with time. The present-day isotopic ratio is therefore a tracer of both formation region and subsequent evolution. Saturn's moon Titan, with its array of photochemically-produced organic molecules, provides a case study for the use of isotope ratios to investigate the evolution of a moon's volatile inventory. ALMA has provided the first measurements of the $^{15}$N/$^{14}$N ratios in HC$_3$N and CH$_3$CN at Titan \citep{Cordiner2018,Iino2020}, and a refined value for the ratio in HCN compared with that obtained by the \textit{Cassini} spacecraft \citep{molter16,vinatier07}. As shown in Figure \ref{fig:15N}, the $^{15}$N/$^{14}$N ratio in these nitriles is significantly larger than in N$_2$, and represents some of the most $^{15}$N-enriched material found anywhere in the Solar System, and similar to that found in $^{15}$N-rich meteoritic ``hot spots'' \citep{bus06}.

This enrichment is explained as a consequence of isotope-selective photodissociation of N$_2$ at high altitudes ($\sim$1000~km) in Titan's thermosphere \citep{dobrijevic18,vuitton19}. The excess of atomic $^{15}$N produced in this way becomes efficiently incorporated into Titan's nitriles \citep[e.g., HC$_3$N and CH$_3$CN;][]{Cordiner2018,nosowitz25} as a result of atmospheric photochemistry, which then mix downward through the atmosphere to become incorporated into progressively larger organic molecules and haze particles, which eventually precipitate out onto the surface. The loss of $^{15}$N into surface organic deposits can cause the bulk atmospheric $^{15}$N/$^{14}$N ratio to decrease \citep{krasnopolsky16}, and the magnitude of this effect can fully account for the difference between cometary values and Titan's present-day (bulk) N$_2$ value \citep{nie10}. Thus, Titan could have accreted its large nitrogen inventory from the same icy reservoir as the comets. 

\subsection{Sulfur}

Jupiter's volcanic moon Io offers another example of a case where an isotopic enrichment traces the long-term evolution of a moon, in this case constraining the longevity of its tidally-driven volcanism. Io hosts a sulfur cycle whereby sulfur-bearing molecules are outgassed via volcanoes and returned to the interior in the form of liquid SO$_2$ (shallow recycling) and sulfates \citep[deep recycling;][]{Hughes2024}. Mass-loss at a rate of 1000--3000 kg/s \citep{Dols2008} takes place from the upper part of Io's atmosphere, which is more abundant in the light isotope due to gravitational stratification \citep{deKleer2024}. Recent ALMA measurements of $^{34}$S in SO and SO$_2$ in Io's atmosphere find a $^{34}$S/$^{32}$S ratio that is 347$\pm$86 \textperthousand{} higher than Solar System average and higher than almost all other known Solar System reservoirs (Figure \ref{fig:15N}). This measurement translates to the loss of 94--99\% of Io's available S inventory over its history. To lose this much S, Io had to have been losing S at a rate of 0.5--5$\times$ its current loss rate for the entire age of the Solar System, indicating that Io has been volcanically active for most or all of its history. As Io's volcanism is driven by tidal heating, which ultimately arises from its Laplace orbital resonance with Europa and Ganymede, this supports a scenario in which the Laplace resonance has been in place for billions of years, as has been proposed by modelers \citep{peale2002}. 

Isotopic ratios measured in comets and protoplanetary disks can provide context for understanding the chemical pathways for the S that gets incorporated into planetary bodies, albeit farther out in the Solar System than Io. Most of the S in disks is present on the surfaces of refractory grains \citep{Riviere-Marichalar2022}, from which isotopic ratios cannot be measured remotely. However, in comet 67P/Churyumov-Gerasimenko, the isotopic ratio of S in grains matches that of gas-phase SO and SO$_2$ and is consistent with Solar System average \citep{Paquette2017}. This suggests that these gas species detected in disks may in fact provide a good tracer of the S in grains, which gets incorporated into rocky planets and moons \citep{Calmonte2017,Altwegg2020}. 
However, where $^{34}$SO$_2$ has been detected in disks (at stellar distances of tens of AU), the $^{34}$S/$^{32}$S ratio is strongly enhanced compared to Solar System, while $^{34}$S/$^{32}$S in SO appears comparable to the Solar System average \citep{booth2024}. The wide variations seen in S isotope ratios between different molecules and disks give it potential as a fingerprint for tracing disk chemistry and the chemical inheritance of planetary bodies, but a greater sample size in both disks and Solar System objects is needed to develop a framework for S inheritance in forming planets.

\section{Outlook}\label{sec:outlook}

A deeper understanding of our Solar System's formation can be gained both by future measurements of Solar System objects themselves and by pushing measurements of disks towards increasingly analogous species and length-scales. Comprehensive isotopic measurements of the CHONS elements across the outer Solar System planets, moons, and small bodies can address key questions regarding volatile inheritance and the long-term evolution of ice and gas reservoirs; while spacecraft missions can make deeper isotopic measurements at individual objects, ALMA can provide measurements of certain species for a wider range of objects, and over longer timescales. 

For placing Solar System bodies in context within the diverse array of protoplanetary disks, the disk measurements should be made at matching stellar distances and of the same molecular species. The exploitation of the longest baselines possible with ALMA grants access to the inner 10 au of most protoplanetary disks \citep{andrews2016}, offering a glimpse of the regions where the planets in our solar system are believed to have formed. These spatial scales are critical to linking the outer disk already extensively studied with ALMA, and the inner few au probed with shorter wavelength observatories like JWST. A more fundamental limitation is that the rocky and icy Solar System bodies accreted their volatile material in the form of ices, which condense out beyond their respective snow lines, while ALMA can only chemically identify gas-phase material. This limitation can be mitigated by observations of molecules sputtered or sublimated from grains. In particular, observations of disks around outbursting stars, which sublimate the ices from grains by temporarily moving the snowline outward, are our best window into the isotopic signature of the ice grains themselves. \cite{tobin2023} measured HDO/H$_2$O in such a disk, and found a ratio comparable to comets as well as to protostellar envelopes (and higher than Earth’s ocean water). This suggests that molecular water is inherited from the protostar through the disk phase and into icy planetesimals without significant chemical alteration occurring, a conclusion that is strengthened by the subsequent finding that the heavier isotopologue D$_2$O is even more enriched in deuterium than HDO \citep{leemker2025}.

There are also now a growing number of detections of embedded, still-forming exoplanets, such as PDS 70 b and c, HD 169142 b and AS 209 b \citep{keppler2018,haffert2019,hammond2023,bae2022}, that provide a unique opportunity to study the immediately-surrounding formation environments for these objects. With sufficiently high angular resolution observations that are able to separate the planet and its circumplanetary disk from the circumstellar disk \citep{benisty2021}, the satellite-forming regions can be extensively characterized, in terms of both their solid and gaseous components. It will be the investment in deep integrations, which permit gas emission to be imaged at these scales, that will lead to the most significant advances in our ability to connect disk observations to volatiles at Solar System planets and moons. For continuum observations, the upcoming Wideband Sensitivity Upgrade \citep{carpenter2023} will enable observations at ALMA’s highest angular resolutions of a larger number of objects in less time, pushing the population of accessible Solar System small body binaries towards increasingly large mass ratios and helping to constrain processes in our own `debris disk' phase.

\clearpage

\noindent \textbf{Conflict of Interest Statement}

The authors have no conflicts of interest to disclose.

\section*{Open Research Section}
Not applicable -- this article does not present new data or software.

\acknowledgments
KdK acknowledges support from the National Science Foundation (NSF) under Grant No. 2238344 through the Faculty Early Career Development Program. MC was supported by (1) a grant from the NASA ROSES Solar System Observations program, (2) NASA’s Planetary Science Division Internal Scientist Funding Program through the Fundamental Laboratory Research work package (FLaRe), and (3) NSF grant AST-2407709.

\bibliography{References}

\end{document}